\documentclass[%
 reprint,
 amsmath,amssymb,
prl,
]{revtex4-1}
\usepackage{graphicx}
\usepackage{caption, subcaption} 
\usepackage{sidecap}
\usepackage{dcolumn}
\usepackage{amsmath}
\usepackage{gensymb}
\usepackage{bm}
\usepackage{float}
\begin{document}

\title{Disordered Mott-Hubbard Physics in Nanoparticle Solids: \\Persistent Gap Across the Disorder-localized--to--Mott-localized Transition}

\author{Davis Unruh$^1$, Alberto Camjayi$^2$, Chase Hansen$^1$, Joel Bobadilla$^2$, Marcelo J. Rozenberg$^3$, and Gergely T. Zimanyi$^1$}
\affiliation{$^1$Physics Department, University of California, Davis, Davis CA 95616}
\affiliation{$^2$Departamento de F\'isica, FCEyN, Universidad de Buenos Aires and IFIBA, Pabell\'on 1, Ciudad Universitaria, 1428 CABA, Argentina}
\affiliation{$^3$Laboratoire de Physique des Solides, CNRS, Université Paris-Sud, 91405, Orsay Cedex, France}

\date{\today}

\begin{abstract}
We show that Nanoparticle (NP) solids are an exciting platform to seek new insights into the disordered Mott-Hubbard physics. We developed a "Hierarchical Nanoparticle Transport Simulator" (HINTS), which builds from localized states to describe the Disorder-localized and Mott-localized phases, and the transitions out of these localized phases. We also studied the interplay between correlations and disorder in the corresponding multi-orbital Hubbard model at and away from integer filling by Dynamical Mean Field Theory. This approach is complementary to HINTS, as it builds from the metallic phase of the NP solid. The mobility scenarios and phase diagrams produced by the two methods are strikingly similar, and account for the mobilities measured in NP solids.
\end{abstract}

\maketitle

Nanoparticle (NP) solids are aggregates of nanometer scale particles with interesting and potentially useful emerging electronic functionalities. In NP solids, the wavefunctions are typically "quantum confined" to the NPs, making their electronic properties tunable with the NP size \cite{doi:10.1021/nn506223h}. This tunability makes them attractive for a wide variety of opto-electronic applications \cite{talapin_prospects_2010}, including photovoltaics \cite{Nozik02, doi:10.1021/jp806791s}, light-emitting diodes \cite{shirasaki2013emergence}, and field-effect transistors (FETs) \cite{Talapin07102005, hetsch_quantum_2013}. However, the same Quantum Confined localization also drives the NP solids insulating, hindering charge transport and thus their utility. Therefore, driving NP solids from their insulating phase across a Metal-Insulator Transition (MIT) into a conducting, metallic phase is a top priority to boost their utility. Recent experimental attempts to cross the MIT included atomic layer deposition (ALD) \cite{liu2013pbse}, substitutional percolation \cite{cargnello2015substitutional}, chemical doping \cite{chen2016metal, choi_bandlike_2012}, and photodoping \cite{talgorn2011unity}. ALD infilling with metal oxides already enhanced mobilities above 7 cm$^2$/Vs \cite{liu2013pbse}. Whether these enhanced-mobility NP solids support coherent metallic transport is still debated. Building on these advances, a few groups developed NP solar cells with 13-16\% power conversion efficiencies \cite{lan201610, jiao12percent_2017, Sanehira_13.4percent}, which is quite promising for their commercial viability.
 
There is a vast literature on the theory of the MIT in disordered systems \cite{Imada}. For NP solids, Ref. \cite{chen2016metal} developed insightful theories. Transport was described as nearest neighbor hopping at high T \cite{yu_variable_2004, liu_mott_2010}, and as Efros-Shklovskii (ES) variable range hopping at low T \cite{efros-shklovskii-76}. The interplay of transport and Coulomb interactions was studied in \cite{PhysRevB.89.235303,chandler_electron_2007}. The concept of the "Coulomb blockade/gap" is sometimes used to refer to the energy cost of the attraction between the hopping electron and the hole it leaves behind,\cite{schmid2006}. This Coulomb blockade suppresses the transport at all fillings. Its dissolution across the MIT was analyzed in granular metals \cite{CoulombGranular,CoulombHopping, GranularSystemsReview}.

The Coulomb blockade is also used to describe an electron on an NP blocking the transport of additional electrons through that NP because of the Coulomb energy cost of double occupancy. This blockade occurs only at integer fillings. It was observed in experiments \cite{dasSarma, Si-transistor, romero,kang_size-_2011}, and analyzed in a few theoretical papers \cite{PhysRevB.89.235303,chandler_electron_2007}, including ours on bilayer NP solids \cite{hansen}. This blockade forces the finite temperature mobility to exhibit minima at integer fillings, exponentially suppressed by the Coulomb gap, and maxima between integer fillings.

Remarkably, in spite of all this progress, profound unmet needs persist in the field. (1) We are not aware of a theory of the MIT in NP solids that starts from the insulating phase. (2) In spite of inviting analogies, the connections between NP solids and the disordered Mott-Hubbard physics have been barely explored \cite{NP-Hubbard1,NP-Hubbard2,dasSarma,NP-Hubbard3}. For example, even though the mobility of NP solids has been observed to exhibit maxima and minima as a function of the filling, these are typically not attributed to Mott-Hubbard commensuration \cite{kang_size-_2011}. 

Working on these challenges promises to pay additional synergistic benefits as well. (a) A theory of the MIT in NP solids that builds out from the insulating side can be very useful for the study of the disordered Hubbard model as well, since that is also lacking a description of the MIT from the insulating phase. And in reverse, the theoretical tools created for the Mott-Hubbard model building from the metallic phase can complement the theory of the MIT in NP solids. (b) There are accepted experimental platforms to study Mott-Hubbard models, such as strongly correlated electron materials, transition-metal oxides \cite{highTc-Dagotto}, especially high temperature superconductors, as well as cold atom systems in optical lattices \cite{christophe-optical}. However, in spite of decades of explorations, basic questions on the physics introduced by disorder effects remain open. Therefore, it can prove to be very helpful to advocate NP solids as a new class of experimental platforms to research the unanswered questions of disordered Mott-Hubbard physics. 

To address the first need of a theoretical method to describe the MIT in NP solids building out from the localized phase, we developed the {\bf HI}erarchical {\bf N}anoparticle {\bf T}ransport {\bf S}imulator HINTS. HINTS is a Kinetic Monte Carlo transport simulator that is extended by an additional metallic transport channel. As such, it is capable of reaching the MIT from the insulating phase. In an initial report we demonstrated that HINTS can indeed reach the MIT \cite{Qu2017}, which we interpreted as a Quantum Percolation Transition. However, we did not connect this MIT to Mott-Hubbard phenomena.

To address the second need, to explore connections between NP solids and Mott-Hubbard physics, we recall that the most widely-studied types of electron localization are the Mott and Anderson mechanisms. Mott localization emerges at $n$=1 electron/site filling, when the local Coulomb repulsion prevents electrons from jumping onto their occupied neighboring sites, thereby localizing electrons. Anderson, or disorder localization takes place in non-interacting systems independent of the filling, where the disorder of the site energies breaks up the site-to-site phase coherence of the originally extended wavefunctions, thereby localizing the electrons\cite{Imada}.

In systems with {\it both} site energy disorder and Coulomb interactions, the Mott and the Anderson localization scenarios interfere, as revealed by studies since the early 80s, most notably by scaling methods \cite{finkelshtein,altshuler,Gabi}. In recent years, this research got re-energized by the adaptation of the Dynamical Mean Field Theory (DMFT)\cite{Vlad1, Vlad2, Vollhardt}. An interesting and surprising prediction of DMFT was that at $n$=1, at intermediate repulsion, the Disorder-localized phase, formed at high disorder, transforms into a metal with decreasing disorder; and further decrease of the disorder transforms the metal into a Mott-insulator. In contrast, at strong repulsion decreasing disorder induces a transition from the Disorder-localized phase directly into the Mott-localized phase without an intervening metallic phase \cite{Vlad2, Vollhardt}. Fillings $n\ne1$ were not investigated. Notably, both the scaling and the DMFT techniques are built with extended wave functions, and thus indicate Anderson/disorder localization only as the boundary where their applicability breaks down. As such, they are ill-suited to describe the Disorder-localized phase itself.

Here are then the main messages of our paper in response of the above needs. (1) NP solids are an exciting experimental and theoretical platform to seek new insights into the disordered Mott-Hubbard physics. (2) HINTS is a powerful method to describe the various transitions starting out from the disorder-localized phase, providing a complementary approach to existing methods building from the  metallic-delocalized phase. (3) For $n$=1 filling and large interactions, HINTS showed that decreasing disorder drives a disorder-localized--to--Mott-localized transition without an intervening metallic phase, i.e. a direct transition, with a persistent gap. (4) For $n\ne1$, HINTS showed that decreasing disorder drives a Disorder-localized--to--Metal transition. (5) We modeled NP solids in their metallic phase as a multi-orbital Hubbard model, which we analyzed with a disordered DMFT technique, complementary to HINTS. This DMFT work at and away from $n=1$ determined how the mobility and the Mott gap were affected by the disorder.

We developed the HINTS code to simulate NP solids by the below-described hierarchical approach. This allowed us to tackle the computational challenges associated with the huge number of degrees of freedom. We computed the electronic energy levels by using a parameterized band structure of individual PbSe NPs with diameters in the 3-8 nm range \cite{kang_electronic_1997}. We then generated a superlattice of NPs with diameters selected from a Gaussian distribution. The evaluation of the electron configuration energies further includes the determination of the on-site Coulomb charging energy $E_c$ with a hybrid empirical-perturbative method \cite{DelerueBook}. The long range part of the Coulomb interaction could be optionally included as the experimental conditions warrant. We modeled the NP-NP transitions as either activated hopping or non-activated metallic transitions, depending on whether the energy difference between the initial and final states was larger or smaller than a hybridization energy. And finally, we computed the mobility of the NP solid by an Extended Kinetic Monte Carlo (KMC) algorithm with a sufficiently small voltage bias. Details of HINTS are provided in the Supplementary Material. We move to presenting the results of the HINTS simulations.
 
\begin{figure}[h!]
    \centering
    \includegraphics[width=8.6cm, keepaspectratio]{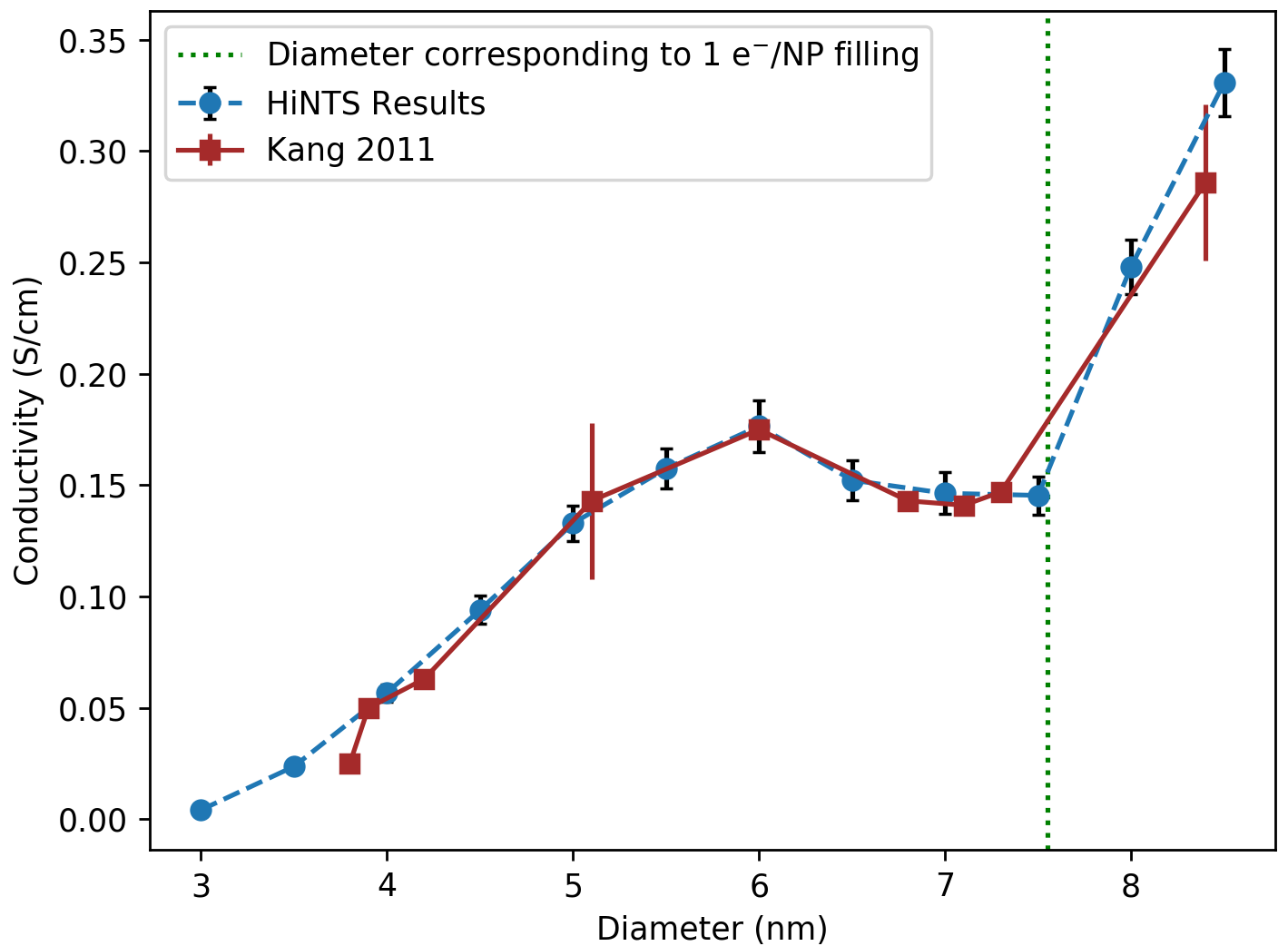}
    \caption{Conductivity v. NP diameter. HINTS simulation overlaid with exp. data. Volumetric electron density .0016 $e/\mathrm{nm}^3$, average charging energy $E_c=100{\rm meV}$, diameter disorder 5$\%$, average hopping energy $t=7{\rm meV}$, $T=200{\rm K}$.}
    \label{fig:MobilityDiameterComparison}
\end{figure}

We first consider the Disorder-localized--to--Mott-localized transition as a function of the filling $n$. Two experimental groups measured the dependence of the mobility on the NP diameter in NP-FETs, and reported an initial rise followed by a maximum  \cite{liu_dependence_2010, kang_size-_2011}. In previous work we simulated NP solids with increasing NP diameters and found that their mobility also exhibited a maximum  \cite{carbone_monte_2013}, consistent with the experiments \cite{liu_dependence_2010}. However, we did not develop an explanation of the results.

We now extend our previous work by using HINTS to compute the conductivity of NP solids as a function of NP diameter well past the maximum. Fig. \ref{fig:MobilityDiameterComparison} shows the HINTS-computed mobilities with increasing NP diameter, overlaid on the experimental data of Kang et al. on PbSe NP FETs \cite{kang_size-_2011}. The HINTS mobilities and the Kang et al. data exhibit remarkable agreement over the extended diameter range. Similar mobilities were reported by the Law group \cite{liu_dependence_2010}.

Notably, the $n$ filling/NP increases with the NP diameter at a fixed volumetric electron density. Conspicuously, the conductivity maximum at a NP diameter of 6 nm is observed to be the beginning of a maximum-minimum-increase pattern, centered on the $n=1$ filling. Interestingly, this is also the behavior expected when scanning across $n=1$ in the repulsive Hubbard model, as the Coulomb repulsion opens a Mott gap at $n=1$, suppressing the transport through the occupied NPs.

In our simulations the temperature dependence of the conductivity is activated for all fillings, having a smaller, disorder-induced gap away from $n=1$, boosted to a larger gap by the Coulomb blockade at $n=1$. Thus, scanning with $n$ through $n=1$ crosses from a Disorder-localized phase through a Mott-localized phase back to the Disorder-localized phase, again as expected in the disordered Hubbard model. These remarkable correspondences between experimental data and our simulations are compelling evidence that NP solids are well-controlled and tunable experimental realizations of the disordered Hubbard model.

Changing the NP diameter also varies other parameters. This convolutes crossing the Mott-localized phase with other trends, such as changing site energies and hopping transition rates. Therefore, next we isolate the mobility's dependence on the filling without impacting the other parameters by analyzing how the mobility of a {\it fixed diameter} NP solid varies when only the filling is tuned. Experimentally this can be achieved by varying the gate voltage applied to a FET formed from an NP solid.
\begin{figure}[h!]
    \centering
    \includegraphics[width=8.6cm, keepaspectratio]{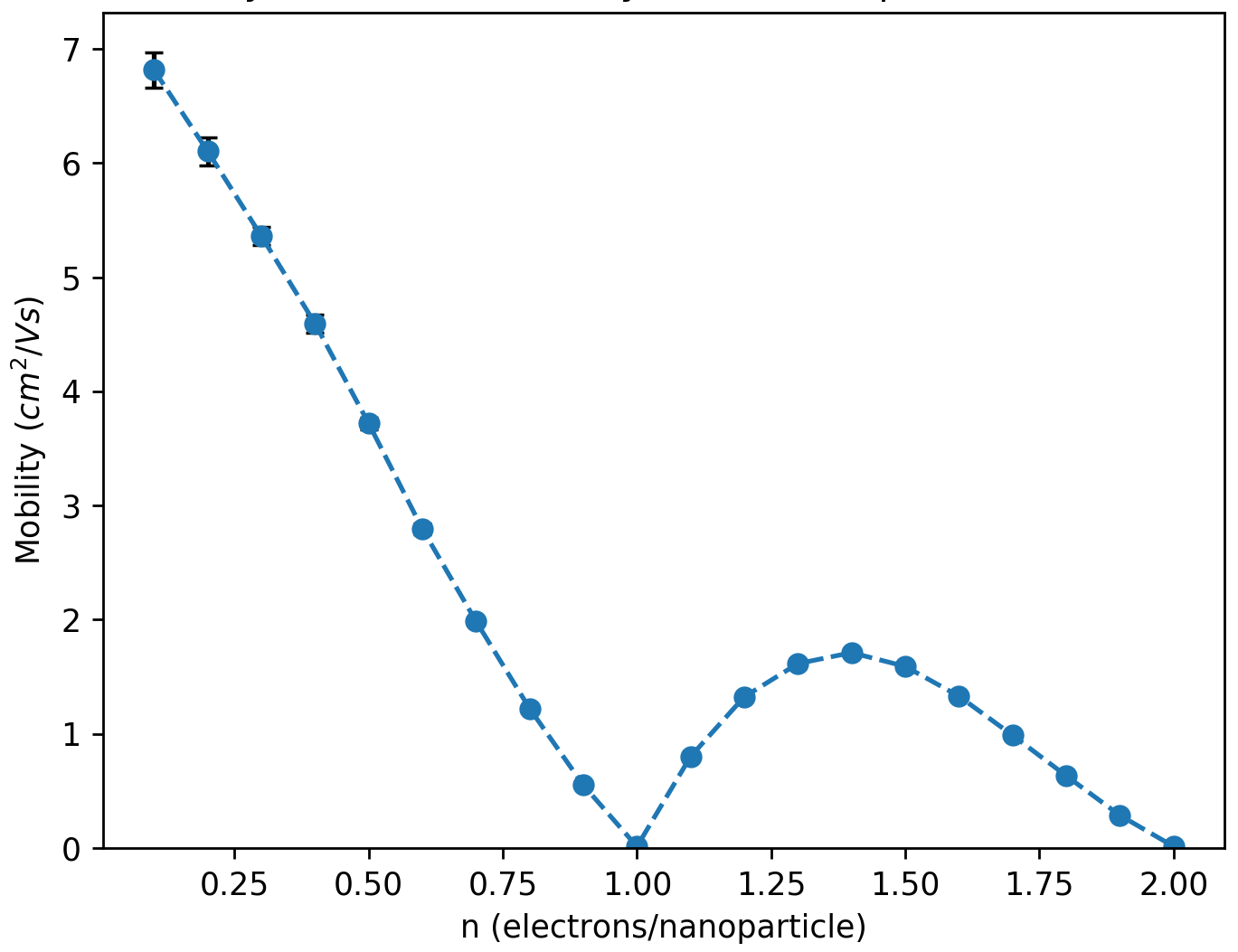}
    \caption{Electron mobility as a function of carrier filling in the disorder-localized Phase with NP diameters of 6.6 $\pm$ 0.3 nm. $T=80 {\rm K}$}
    \label{fig:Mobility}
\end{figure}

Fig. \ref{fig:Mobility} shows the HINTS-simulated mobility of the NP solid as the filling $n$ was varied. As before, the mobility exhibits pronounced minima at integer fillings, accompanied by maxima close to half-integer fillings. This is a re-confirmation of the Disorder-localized-to-Mott-localized transition in the NP solids as the filling $n$ is tuned.

\begin{figure}[h!]
    \centering
    \includegraphics[width=8.6cm, keepaspectratio]{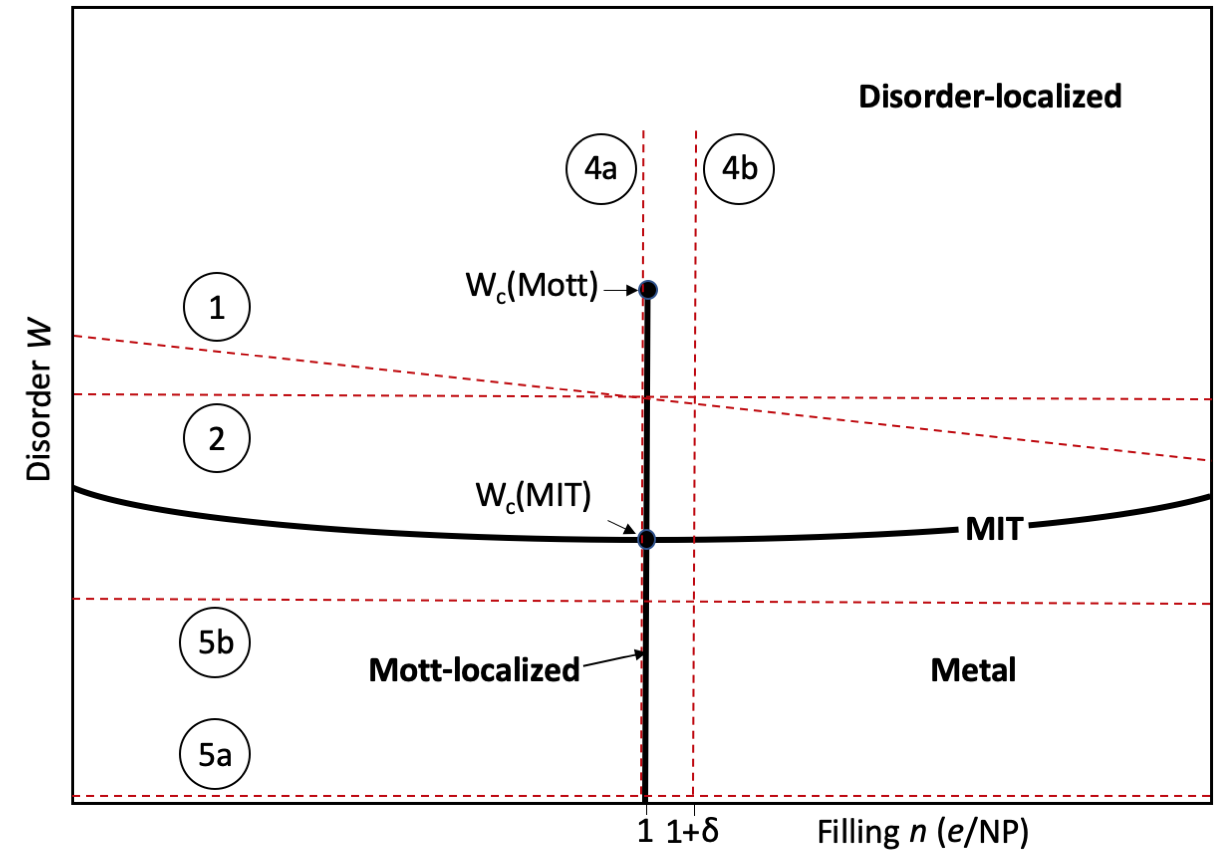}
    \caption{Qualitative phase diagram of NP solids in the (disorder \textit{W}--filling \textit{n}) space, for $U>>W$. The red dashed lines, labeled by circled numbers indicate the reported scans.}
    \label{fig:PhaseDiagram}
\end{figure}

Fig. \ref{fig:PhaseDiagram} provides a schematic guide for the various scans performed across the phase diagram. In scans 1 and 2, the filling was tuned across $n=1$, respectively shown in Figs. 1 and 2. Scan 1 is tilted to indicate the simultaneous change of the other parameters.

\begin{figure}[h!]
    \centering
    \includegraphics[width=8.6cm, keepaspectratio]{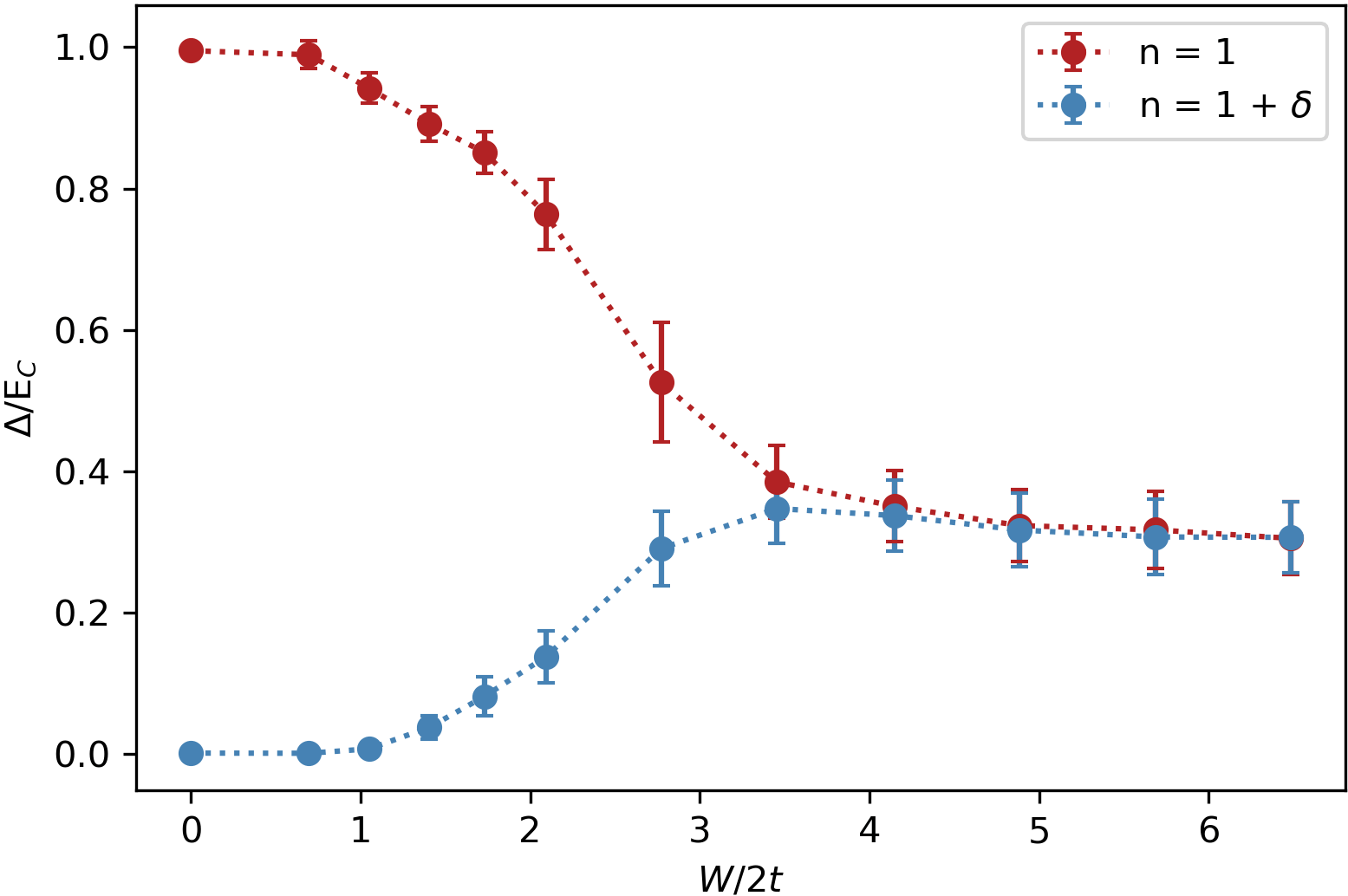}
    \caption{The disorder dependent gap $\Delta$ (W): (a) $n=1$; (b) $n=1+\delta$. $W_c(Mott)/2t\approx 3.5$; $W_c(MIT)/2t\approx 1.5$.}
    \label{fig:GapProgression.png}
\end{figure}

Next, we analyze the Disorder-localized--to--Mott-localized transition as a function of disorder $W$ around $n$=1. Fig. \ref{fig:GapProgression.png} shows the $W$ dependent gap $\Delta(W)$ at fillings $n$=1 (scan 4a), and $n=1+\delta$, $\delta=0.001$ off the integer $n$=1 (scan 4b). At $n$=1 and small disorder the gap $\Delta$ is Mott-like, with a value set by the charging energy $E_c$ that suppresses transport, thus creating a Mott-localized phase. With increasing disorder, the gap $\Delta$ gets renormalized to a lower value by the disorder. The physics of this lower gap region is clarified by scan 4b of $\Delta(W)$ at a filling $n=1+\delta$. The gap $\Delta(W)$ is the same for $n=1$ and $n=1+\delta$ above a critical value $W_c(Mott)/2t\approx 3.5$. The gap's insensitivity to the charging energy and to commensuration identifies this phase as a Disorder-localized phase, into which the Mott-localized phase of $n$=1 transitions for $W>W_c(Mott)$. The relative constancy of $\Delta (W)$ in this phase is explained by the charging energy $E_c$ screening the disorder $W$. Clearly, the transition at $n=1$ across $W_c(Mott)$ is a direct one, having a persistent gap and no intervening gapless metallic phase.

A qualitatively similar direct Mott-to-Anderson transition was found by DMFT with a disorder scan in the Hubbard model at $n$=1 for $U/t>10$ \cite{Vlad2}. Thus, our work and Ref.\cite{Vlad2} both report a direct Mott-localized--to--Disorder-localized transition with a persistent gap as $W$ crosses a $W_c(Mott)$ in the disordered Hubbard model at $n$=1. Next, we go beyond previous work by exploring the filling dependence away from $n=1$.

Scan 4b at $n=1+\delta$ shows that as the disorder $W$ decreases, the gap shrinks to zero at $W_c(MIT)/2t \approx 1.5$. HINTS does not track the phases of the electron wavefunctions, and thus describes the transport with decreasing disorder only to $W=W_c(MIT)$. We identify $W_c(MIT)$ as the Metal-Insulator Transition of the $incommensurate$ NP solid from the Disorder-localized phase to a metal. This MIT at $n=1+\delta$ is disorder-driven, as opposed to the interaction-driven Mott transition at $W=W_c(Mott)$ at $n=1$.

Our results also mean that a filling scan across $n=1$ at a fixed disorder $W$ between $W_c(MIT)$ and $W_c(Mott)$ induces a filling-driven Disorder-localized--to--Mott-localized transition with a persistent gap, another new result beyond the scope of previous works \cite{Vlad2}.  

We now turn to the Disordered-metal--to--Mott-localized transition as function of filling $n$. We already established that the physics of the NP solids in the localized phases is analogous to the disordered Hubbard model. Now we complete the analogy by describing the metallic phase of the NP solid as a Hubbard model as well whose sites represent the individual NPs. We put the NP solid on a regular lattice to start from a bona fide metallic state and explore the combined effect of Coulomb repulsion and site energy disorder. This approach is complementary to HINTS, which starts from localized states. We use similar parameters as before by associating the on-site Hubbard repulsion $U$ with the NP charging energy to take $U=E_c=100 meV$, and the kinetic term with the NP-NP hopping amplitude of $t=7meV$, making $U/2t = 7$. We take into account the degeneracy of the electronic states of the PbSe NPs \cite{PbSe} by adopting a 8-fold degenerate multi-orbital Hubbard model \cite{mjr-degHubbard}.

Since $t/U<<1$, the Hubbard model is in strong coupling. Therefore, we adopt the  Dynamical Mean Field Theory (DMFT)\cite{DMFTrev}. DMFT proved very successful for the study of the interplay between correlation effects and the Mott-localization of an electron band, producing many useful insights into the Mott metal-insulator transition. Crucially, DMFT can also be extended to include disorder effects \cite{Vlad1, Vlad2, Vollhardt} (see Sup. Mat. for details).

The DMFT approach maps the original lattice model onto an auxiliary quantum impurity model supplemented with a self-consistency condition. The quantum impurity problem is then tackled by numerical simulations that we performed using the continuous-time quantum Monte Carlo (CTQMC) method\cite{CTQMCrev}, described in Ref. \cite{CTQMC_code}. That method samples a diagrammatic expansion of the partition function in powers of the impurity-bath hybridization. For simplicity, for the non-interacting electrons we adopt a semi-circular density of states of bandwidth $4t$. The disorder is introduced through the site energy random variable (diagonal disorder) with a uniform distribution in the energy interval $[-W,W]$. (see Sup. Mat. for details.)
\begin{figure}[h!]
    \centering
    \includegraphics[width=8.6cm, keepaspectratio]{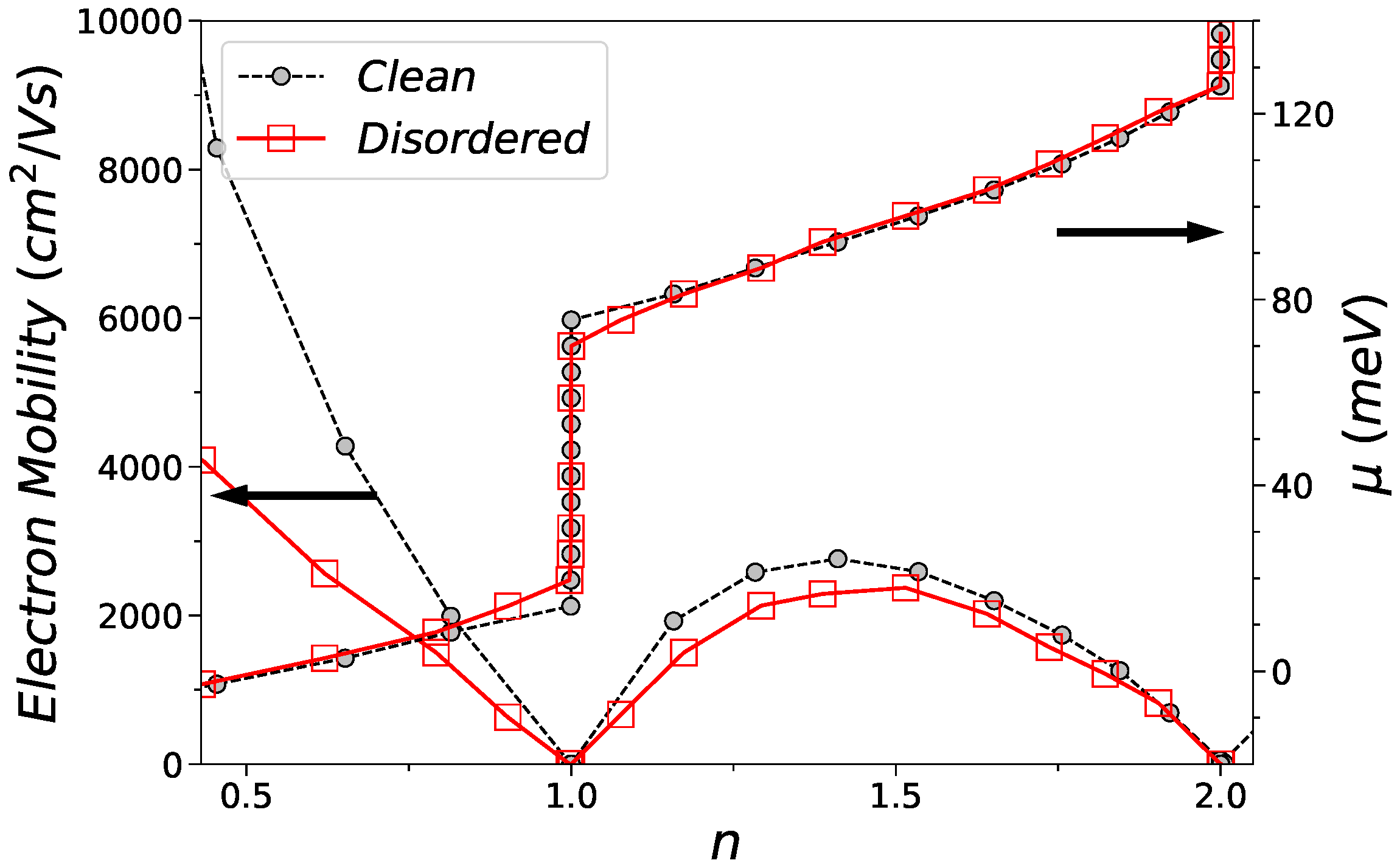}
    \caption{Mobility and chemical potential $\mu$ as filling scans from the Metallic Phase through the Mott gap.} \label{fig:MobilityDensityInsulator}
\end{figure}

Fig. \ref{fig:MobilityDensityInsulator} shows the electron mobility and the chemical potential $\mu$ of this Hubbard model as function of the filling $n$ for the clean and disordered cases (scans 5a/5b in Fig.\ref{fig:PhaseDiagram}). In the clean case, $\mu$ shows a jump at $n=1$: this indicates the emergence of a Mott gap that localizes the electrons\cite{mjr-degHubbard}. The Mott gap makes the mobility exhibit a minimum as the filling crosses $n$=1 (i.e. band filling=1/8, well below half filling). Clearly, DMFT established that the Mott gap/Coulomb blockade also suppresses the mobility in the metallic phase when the filling is scanned across $n=1$, just like in the insulating phase (cf. Fig. \ref{fig:Mobility}). This confirms and completes the phase diagram obtained by HINTS that was built from the localized phase.

A main result of the DMFT study is that the clean trends persist even for a substantial disorder $W/2t=1$. Indeed, the Mott gap (jump of $\mu$) at $n=1$ only decreases by a small amount and remains robust. This is reasonable since $W/2t=1<<W_c(Mott)/2t=3.5$. At the same time, the mobility away from $n=1$ is reduced notably because the $W/2t=1$ disorder is relatively close to the off-commensuration MIT at $W_c(MIT)/2t=1.5$. Finally, Fig. \ref{fig:MobilityDensityInsulator} shows that the relative reduction of the mobility by the disorder is strongest at small $n$. This makes physical sense, since at small $n$ the Fermi energy becomes comparable to the disorder, and thus the relative importance of the disorder grows. To our knowledge, the present study of disorder-DMFT is the first one done for a multi-orbital model as a function of electron filling. 

One of the most appealing results of our study is that despite HINTS and disorder-DMFT describing the MIT from dual opposite starting points, they produce qualitatively the same scenarios. This is well illustrated by the strikingly similar Figs.\ref{fig:Mobility} and \ref{fig:MobilityDensityInsulator}. The discrepancy between the magnitudes of the insulating and metallic mobilities is due to the insulating phase having a gap that exponentially suppresses the mobility, whereas the metallic phase lacking this suppression factor. Reassuringly, experiments often report orders of magnitude changes when the mobility crosses the MIT\cite{Imada}.

To conclude, we argued that NP solids are an exciting platform to seek new insights into the disordered Mott-Hubbard physics. We presented HINTS as a new powerful method to describe the various transitions from the Disorder-localized phase; and showed how the DMFT method can complement these studies from the delocalized phase. In particular, at $n=1$ filling, HINTS showed that at large interactions the Disorder-localized--to--Mott-localized transition occurs with a persistent gap. These results articulate an important message for how to optimize the mobility of NP-based optoelectronic 
and solar devices: tune the disorder below the Metal-Insulator transition, and tune the electron filling far from commensurate values. In FETs this can be done by tuning the gate voltage. In solar cells, by cobaltocene doping\cite{DopingPbSe,DopingPbSeMarton}.

Acknowledgments: We acknowledge helpful discussions with L. Qu, M. Voros and M. Law. This work was supported by the UC Office of the President under the UC Laboratory Fees Research Program Collaborative Research and Training Award LFR-17-477148. A.C. and J.B. gratefully acknowledge
support from CONICET and UBACyT.


%

\end{document}


\title{Disordered Mott-Hubbard Physics in Nanoparticle Solids: \\Persistent Gap Across the Disorder-localized--to--Mott-localized Transition: Supplemental Material}

\author{Davis Unruh$^1$, Alberto Camjayi$^2$, Chase Hansen$^1$, Joel Bobadilla$^2$, Marcelo Rozenberg$^3$, and Gergely T. Zimanyi$^1$}
\affiliation{$^1$Physics Department, University of California, Davis, Davis CA 95616}
\affiliation{$^2$Departamento de F\'isica, FCEyN, Universidad de Buenos Aires and IFIBA, Pabell\'on 1, Ciudad Universitaria, 1428 CABA, Argentina}
\affiliation{$^3$Laboratoire de Physique des Solides, CNRS, Université Paris-Sud, 91405, Orsay Cedex, France}

\date{\today}

\maketitle
\section{HiNTS numerical calculation details}
\underline{Simulation Samples}: The NP samples used for our simulations consist of a thin film geometry, typically containing 800 PbSe nanoparticles arranged as a 20x2x20 ($x, y, z$) cubic array. This is appropriate for Field Effect Transistor (FET) NP solids where transport primarily occurs in the top few layers, with small penetration depths. 

Each NP's diameter has a standard deviation $\sigma$, taken as the fixed value 0.3 nm. This is representative of corresponding experiments \cite{LawExperiment}. The mean inter-NP distance is taken as twice the average ligand length in the system, with variation due to the aforementioned size disorder. The ligand length was set to 0.5 nm to correspond to EDT (1,2-Ethanedithiol), which capped the nanoparticles in most experiments which do not include necking.

Transport is induced in the $\hat{z}$ direction via the application of a small voltage bias. This voltage is kept small to ensure that our samples remained in the linear regime of the I-V curve. Further, disorder of nanoparticle energies does not average exactly to zero, creating an internal bias. To remove this, the voltage bias is applied both in the forward and reverse transport directions (2 different simulation runs), and the resulting current is pairwise averaged. 

\underline{Electronic NP Structure}: The HOMO and LUMO band energies of the NPs are calculated using the $k\cdot p$ method of Kang \& Wise \cite{KangWise}. We apply a rigid shift to align the infinite diameter limit of the conduction band edge with the bulk work function of PbSe. 

On-site Coulombic repulsion $E_C$, also referred to as the on-site charging energy, is the energy cost that needs to be paid to load each electron onto a given nanoparticle. In general, the accounting is defined by:
\begin{equation}
    E_C = n(\Sigma^0 + \frac{(n-1)}{2}\Sigma)
\end{equation}
where $n$ is the number of electrons that will be on the NP after the electron is added. $\Sigma^0$ is the energy that needs to be paid upon the load of the first charge onto the neutral NP, while $\Sigma$ is the extra energy it takes to load each additional charge (due to interactions with other charges as well as their induced image charge). 

In our model, $E_C$ is calculated using the single NP empirical-perturbative hybrid calculations of Delerue \cite{Delerue}, consistent with single NP \textit{ab initio} calculations \cite{zunger} and the self-capacitive term used by experimental reports \cite{LawExperiment}. This method gives the results:
\begin{equation}
    \Sigma^0 = \frac{q^2}{8\pi \epsilon_0 R}(\frac{1}{\epsilon_{\mathrm{solid}}} - \frac{1}{\epsilon_{\mathrm{NP}}}) + 0.47\frac{q^2}{4\pi \epsilon_0 \epsilon_{\mathrm{NP}} R}(\frac{\epsilon_{\mathrm{NP}} - \epsilon_{\mathrm{solid}}}{\epsilon_{\mathrm{NP}} + \epsilon_{\mathrm{solid}}})
\end{equation}
and
\begin{equation}
    \Sigma = \frac{q^2}{4\pi \epsilon_0 R}(\frac{1}{\epsilon_{\mathrm{solid}}} + 0.79\frac{1}{\epsilon_{\mathrm{NP}}})
\end{equation}
For the dielectric constant inside the NP, we assume that it will be the bulk dielectric constant of PbSe, taken to be 22.0. To properly treat the dielectric constant of the medium, we account for both the organic ligand shell of the NP, as well as the presence of neighboring NPs. We assume that the ligands themselves have a dielectric constant of 2.0. The dielectric constant of the entire solid is then calculated using the Maxwell-Garnett (MG) effective medium approximation:
\begin{equation}
    \epsilon_{\mathrm{solid}} = \epsilon_{\mathrm{ligand}}\frac{\epsilon_{\mathrm{NP}}(1 + \kappa f) - \epsilon_{\mathrm{ligand}}(\kappa f - \kappa)}{\epsilon_{\mathrm{ligand}}(\kappa + f) + \epsilon_{\mathrm{NP}}(1-f)}
\end{equation}
where $\kappa$ is 2 for spherical NPs, and f is the filling factor.

\underline{Kinetic Monte Carlo Framework}: Monte Carlo (MC) algorithms are a class of numerical algorithms which incorporate random numbers in attempts to accurately simulate real-world problems. Broadly, this is implemented by calculating the probability of specific events, and then executing them if a random number between 0 and 1 falls in that specified range.

A Kinetic Monte Carlo algorithm is a specific MC which seeks to accurately simulate the evolution in time of physical systems. It does so by propagating the physical system forwards step by step, executing a single event and evolving time accordingly at every step. In this manner, KMC simulations mimic explicitly the evolution of real-world systems. 

Our Kinetic Monte Carlo simulator is implemented according to the BKL algorithm. At each step, the KMC tabulates the probabilities of all possible events occurring, and then chooses which one to execute by choosing a random number $r_1$ between 0 and 1. Time is then evolved by choosing another random number $r_2$ (also between 0 and 1). The specifics are as follows.

Once the simulation is initialized (and all charges have been placed at random into the network), the time-evolution starts by determining the transition rates $\Gamma$ of all possible events, which will be locally updated after each algorithm step. In each step, the event $j$ is executed using the random number $r_1$ for which the following equation is satisfied:
\begin{equation}
    \sum^{j-1}_{i=1}\Gamma_i < r_1\Gamma_{\textrm{sum}} < \sum^{N}_{i = j + 1}\Gamma_i
\end{equation}
where
\begin{equation}
        \Gamma_{\textrm{sum}} = \sum^{N}_{i = j + 1}\Gamma_i
\end{equation}

We have adopted Miller-Abrahams single-phonon assisted hopping as our framework for calculating these transition probabilities. An alternate approach also included as an option in our code is the Marcus theory of multi-phonon activated transitions, but Miller-Abrahams typically yields the more accurate results. Under Miller-Abrahams, the probabilities are given by:
\begin{equation}
    \Gamma_{i \rightarrow j} = \nu\beta_{ij}\exp{(\frac{-\Delta E_{ij}}{k_b T})}
\end{equation}
where the attempt frequency $\nu$ is chosen to match experimental data. $\Delta E_{ij}$ can be calculated as $\Delta E_{ij} = \Delta E_{ij}^{sp} + \Delta E_{ij}^{F} + \Delta E_{ij}^{C}$. $E^{sp}$ is the single-particle band energy. $E^C$ is the charging energy. Finally, $\Delta_{ij} E^F$ is the energy contribution from the external electric field, $\Delta_{ij} E^F = q\frac{V}{L_z}(z_j - z_i)$, where $L_z$ is the entire length of the sample in the $\hat{z}$ direction, and $(z_j - z_i)$ is the z distance between the two NPs. 

Here it should also be noted that $\beta_{ij}$ is the tunneling amplitude as given by the WKB approximation:
\begin{equation}
    \beta_{ij} = \exp{(-2\Delta x\sqrt{\frac{2 m^{*} (E_{vac} - E_{tunneling})}{\hbar^2}})}
\end{equation}
where the energy of the well that the electron resides in, $E_{tunneling}$, is taken to be $\frac{E_i + E_j}{2}$, the average of the initial and final energy states. This is the approach of Chandler \& Nelson \cite{ChandlerNelson}. 

For the calculations of Fig. 4, an additional non-activated metallic transport channel was included, where $\Gamma_{i \rightarrow j} = \nu\beta_{ij}$. Transitions via this channel are allowed if $\Delta E_{ij}$ is less than an NP-NP hybridization energy $E_H$. $E_H$ depends on the electron wavefunction overlap between NPs $i$ and $j$, and for simplicity is taken to be approximately equal to the half bandwidth $W$.

After this execution, time is evolved according to the second random number, $r_2$:
\begin{equation}
    \Delta t = -\frac{\ln{r_2}}{\sum_{ij}\Gamma_{j\rightarrow i}}
\end{equation}

Finally, the list of all possible events is updated to reflect the current state of the system, and the rate table is updated accordingly. 

Our simulations typically run for 500,000 events, well into the steady state. Data is only collected after the transients have dissipated.



Mobility is measured in our simulation as:
\begin{equation}
    \mu = \frac{(\mathrm{harvested \quad charge})*L}{\mathrm{time}*N*F_{ext}}
\end{equation}
with $L$ the length of the simulation box, $N$ the total number of carriers, and $F_{ext} = eV/L$. The harvested charge is the total charge of the electrons which have reached the drain electrode. Our simulations use periodic boundary conditions, so these electrons will pass through the drain electrode to the source electrode to be reintroduced to the sample. Any electron that travels in the opposite direction gives a negative addition to the current.

Conductivity can be taken directly from mobility. The classical relation is:
\begin{equation}
    \sigma = ne\mu
\end{equation}
where $n$ is the volumetric electron density $N/V$.

\section{DMFT numerical calculation details}

We study a four orbital (labeled by $a,b=1,2,3,4$) Hubbard model with diagonal disorder. The Hamiltonian reads:

\begin{equation}
H = \sum_{\langle i,j\rangle,a,\sigma} t_{ij} d_{i,a,\sigma}^\dagger d_{j,a,\sigma} + \sum_{i,a,\sigma} (w_i - \mu) n_{i,a,\sigma} + \sum_i H^\mathrm{int}_i.
\end{equation}
Here $i,j$ label sites in a lattice and $\langle i,j \rangle$ limit the summation over nearest neighbors only, 
$n_{i,a,\sigma} = d_{i,a,\sigma}^\dagger d_{i,a,\sigma}$ is the density of electrons of spin $\sigma$ 
in orbital $a$ on site $i$, $\mu$ is the chemical potential, and $t_{ij}$ is the (nearest neighbor) hopping. 
The disorder is introduced
through the (quenched) random site potential energy $w_i$, 
which for simplicity
we assume to be independent of orbital and spin index.

For the interaction term, we take the simplest of the standard Slater-Kanamori form

\begin{equation}
H^\mathrm{int} = \sum_a U n_{a,\uparrow} n_{a,\downarrow} + \sum_{a \neq b,\sigma,\sigma^\prime} U n_{a,\sigma} n_{b,\sigma^\prime}
\label{int}
\end{equation}
where $U$ is the density-density Coulomb interaction. 
Inter-orbital and Hund's couplings are usually included 
in the study of materials, where the physics is dominated by electrons in the transition-metal t$_\mathrm{2g}$ orbitals.
The model without disorder has been extensively studied \cite{lombardo,werner,kita,luca,xidai,antipov} in that context, and the role of 
Hund's coupling and rotational invariance has been highlighted. 
For the NP solid, spin-orbit interactions are not important. 
Only the 8-fold degeneracy (4 orbitals plus spin) 
and density-density Coulomb repulsion are relevant.

We solve the model using the single-site dynamical mean-field approximation (DMFT), which neglects the momentum 
dependence of the self-energy and where the local single-particle Green function is determined self-consistently \cite{george}.
If in this approach the effect of local disorder is taken into account through the arithmetic mean of the local 
density of states, one obtains, in the absence of interactions, the well known coherent potential approximation (CPA). This method is appropriate for our current purpose, namely, study the effect of disorder on the mobility of the correlated metal state as function of the occupation, and its effect of the gap at $n$=1.
The CPA-DMFT method, however, does not capture the physics of Anderson localization and other disorder-DMFT approaches have been proposed. For instance, ``statistical DMFT,'' where not only 
the averages but the probability distribution itself are 
determined self-consistently, or ``Typical Medium Theory'' 
where geometric averages are considered \cite{dis1,vlad,dis2,vlad2,aguiar,radonjic,byczuk,dis3}.
These methods are numerically more costly and more relevant to
the specific study of the Mott-Anderson transition. The extension
of our work in those directions is an interesting perspective.

Here we will consider quenched, uncorrelated, local site energies $w_i$ drawn from a probability distribution function given by:
\begin{equation}
P(w_i) = \frac{1}{2W}\Theta(W-|w_i|)
\end{equation}
with $\Theta$ the Heaviside step function. We will set the disorder strength $W \ll U$ where CPA is a good approximation. 

In DMFT for disordered electrons, a quasiparticle is characterized by a local but site dependent self-energy $\Sigma_i(i\omega_n)$, with $\omega_n$ the fermionic Matsubara frequencies. To obtain those self-energies the problem is mapped into an ensemble of Anderson impurity problems embedded in a self-consistently calculated conduction bath. 
Here we adopt a semicircular density of state (DOS) so, in this particular case, the hybridization function is given by:
\begin{equation}
    \Delta(i\omega_n)=t^2 G_\mathrm{avg}(i\omega_n)
\end{equation}
and the average local Green's function, $G_\mathrm{avg}$, is obtained from the self-consistency condition:
\begin{equation}
    G_\mathrm{avg}(i\omega_n)=\left\langle\frac{1}{i\omega_n+\mu-w_i-\Delta(i\omega_n)-\Sigma_i(i\omega_n)}\right\rangle
\end{equation}
where $\langle\dots\rangle$ stands for the arithmetic average over the distributions of $w_i$. Since we are dealing with a rather demanding 4-orbital model, we take ten random values of site energies at each self-consistency loop.

The simulations were performed using the continuous-time quantum Monte Carlo implemented in 
Refs.~\cite{haule}, which samples a diagrammatic expansion of the partition function in powers of the impurity-bath 
hybridization \cite{gull}.

To calculate the mobility we obtain the dc conductivity $\sigma$ from DMFT:
\begin{equation}
    \sigma=\sigma_0\int_{-\infty}^\infty \mathrm{d}\varepsilon \int_{-\infty}^\infty \mathrm{d}\omega D(\varepsilon) \rho(\omega,\varepsilon)^2 \left( -\frac{\partial f}{\partial\omega}\right)
\end{equation}
where $\sigma_0$ is a constant, $f$ the Fermi function, $D(\varepsilon)$ is the DOS,  $\rho(\omega,\varepsilon)=-\frac{1}{\pi}G(i\omega_n \to \omega+i0^+,\varepsilon)$ is the spectral function and $G(i\omega_n,\varepsilon)$ is given by:
\begin{equation}
G(i\omega_n,\varepsilon)=\frac{1}{i\omega_n+\mu-\varepsilon-\Sigma_\mathrm{avg}(i\omega_n)}
\end{equation}
with $\Sigma_\mathrm{avg}$ obtained from the Dyson equation:
\begin{equation}
    \Sigma_\mathrm{avg}(i\omega_n)=i\omega_n+\mu-t^2G_\mathrm{avg}(i\omega_n)-\left[G_\mathrm{avg}(i\omega_n)\right]^{-1}.
\end{equation}

In the calculation we set the units of energy such that $t=0.5$, $U=7$ and $W=1$.
